%% file: main.tex
\newif\ifcomm
\commfalse

\newif\ifs
\sfalse

\newif\ifblind
\blindfalse

\newif\ifhyperref
\hyperreftrue

\newif\ifconf
\conftrue

\newcounter{format}
\newif\ifacm
\ifnum \value{format} < 2 \acmtrue \else \acmfalse \fi
\newif\ifacmart %
\ifnum \value{format} = 0 \acmarttrue \else \acmartfalse \fi
\newif\ifusenix
\ifnum \value{format} = 2 \usenixtrue \else \usenixfalse \fi
\newif\ifhotnets
\ifnum \value{format} = 3 \hotnetstrue \else \hotnetsfalse \fi
\newif\ifieee
\ifnum \value{format} = 4 \ieeetrue \else \ieeefalse \fi
\newif\ifnines
\ifnum \value{format} = 5 \ninestrue \else \ninesfalse \fi

\ifconf
    \newcommand{\Conf}[1]{#1}
    \newcommand{\TR}[1]{}
    \newcommand{\Journal}[1]{}  %
    \newcommand{\OnlyTR}[1]{}   %
\else
    \newcommand{\Conf}[1]{}
    \newcommand{\TR}[1]{#1}
    \newcommand{\Journal}[1]{#1}  %
    \newcommand{\OnlyTR}[1]{}   %
\fi

\ifblind
    \documentclass[sigconf, 10pt, anonymous, nonacm]{acmart}
\else
    \documentclass[sigconf, 10pt, nonacm]{acmart}
\fi

\ifieee %
    \usepackage[noadjust]{cite} %
	\usepackage[cmex10]{amsmath}
	\usepackage{amsthm}  %
    \newtheoremstyle{boldthm}{}{}{\itshape}{}{\bfseries}{.}{ }{\thmname{#1}\thmnumber{ #2}\thmnote{ (#3)}} %
    \theoremstyle{boldthm}
            
\else
    \usepackage{amsmath}
    \usepackage{amsthm}

\fi

\ifnines
    \usepackage{graphicx}
    \usepackage[tt=false, type1=true]{libertine}
    \usepackage{mathptmx}
    \usepackage{amssymb}
    \usepackage[varqu]{zi4}
    \usepackage[T1]{fontenc}
    \usepackage[font={footnotesize},labelfont={footnotesize,bf},textfont={footnotesize,it}]{caption}
\fi
\ifhotnets
    \usepackage{times}  
\fi

\ifacmart 
\else 
    \ifusenix
    \else
        \ifhotnets 
        \else
            \ifhyperref
                \usepackage[hidelinks]{hyperref} %
                \hypersetup{pdfstartview=FitH,pdfpagelayout=SinglePage}
            \fi
        \fi
    \fi
\fi

\ifacmart
\else

\fi

\ifieee %
    \newcommand{\bp}{\begin{IEEEproof}}     %
    \newcommand{\bpo}{ \begin{IEEEproof}[Proof Outline] }
    \newcommand{\ep}{\end{IEEEproof}}       %
    \newcommand{\proofof}[1]{\begin{IEEEproof}[Proof of #1]} %
\else
    \newcommand{\bp}{\begin{proof}}
    \newcommand{\bpo}{ \begin{proof}[Proof Outline] }
    \newcommand{\ep}{\end{proof}}       %
    \newcommand{\proofof}[1]{\begin{proof}[Proof of #1]} %
\fi

\ifieee
    \usepackage[justification=justified, font=footnotesize]{caption}
    \usepackage[labelformat=simple, font=footnotesize]{subcaption}  
    \DeclareCaptionLabelSeparator{periodspace}{.\quad}
    \ifconf
    \else
    \fi
\else %
    \ifnines \else
        \ifs %
            \usepackage[small]{caption} 
        \else
            \usepackage{caption}
        \fi
    \fi %
    \usepackage[labelformat=simple]{subcaption}
\fi

\usepackage{graphicx}
\usepackage{xurl}
\usepackage[normalem]{ulem} %
\usepackage{xcolor} %

\usepackage{xspace}
\usepackage{balance}
\usepackage{booktabs}  %
	
\usepackage{array}
	\newcolumntype{C}[1]{>{\centering\let\newline\\\arraybackslash\hspace{0pt}}m{#1}}

\usepackage[inline,shortlabels]{enumitem} %
\newcommand{\bl}{\begin{enumerate*}[(1)]}
\newcommand{\el}{\end{enumerate*}}

\usepackage[capitalize,noabbrev]{cleveref}
\crefname{equation}{Eq.}{Eqs.}
\Crefname{equation}{Eq.}{Eqs.}
\crefname{figure}{Fig.}{Figs.}
\Crefname{figure}{Fig.}{Figs.}
\crefname{table}{Table}{Tables}
\Crefname{table}{Table}{Tables}
\crefname{property}{Property}{Properties}
\Crefname{property}{Property}{Properties}
\crefformat{section}{\S#2#1#3}
\crefformat{subsection}{\S#2#1#3}
\Crefformat{section}{\S\S#2#1#3} %
\Crefformat{subsection}{\S\S#2#1#3} %

\ifcomm
  \newcommand{\mycomm}[3]{{\footnotesize{{\color{#2} \textbf{[#1: #3]}}}}}
  \newcommand{\Fmycomm}[3]{{\color{red} \footnote{{{\color{#2} \textbf{[#1: #3]}}} }}}
\else
    \newcommand{\mycomm}[3]{}
    \newcommand{\Fmycomm}[3]{}
\fi

\ifs
	\usepackage[compact]{titlesec} %

    \setlist{leftmargin=*} %
    \newcommand{\T}[1]{\par\vspace{2pt plus 1pt minus 1pt}\noindent\textbf{#1}} %
    \newcommand{\Ts}[1]{\par\noindent\textit{#1}} %

\else

    \newcommand{\T}[1]{\par\smallskip\noindent\textbf{#1}} %
    \newcommand{\Ts}[1]{\par\smallskip\noindent\textit{#1}} %
\fi
\newcommand{\mypar}[1]{\T{#1.\xspace}}
\newcommand{\mypars}[1]{\Ts{#1.\xspace}}

\newcommand{\be}{\begin{equation}}
\newcommand{\ee}{\end{equation}}

\newcommand{\unit}[1]{\;\mathrm{#1}} %

\newcommand{\vx}{\checkmark\kern-1.1ex\raisebox{.7ex}{\rotatebox[origin=c]{125}{--}}} %

\usepackage{pifont} %

\providecommand{\ie}{{i.e.,}\xspace}
\providecommand{\eg}{{e.g.,}\xspace}

\newcommand{\newVar}[2]{\newcommand{#1}{\ensuremath{#2}\xspace}}
\newcommand{\renewVar}[2]{\renewcommand{#1}{\ensuremath{#2}\xspace}}
  \newVar{\server}{S}
  \newVar{\client}{C}
  \newVar{\rclient}{R_c}
  \newVar{\rserver}{R_s}
  \renewVar{\th}{\ensuremath{^\text{th}}}

\begin{document}

\title{Incast-Free MoE Rate-Based Scheduling}

\ifblind
\else
    \ifhyperref
        \newcommand{\aut}[2]{#1\texorpdfstring{$^{#2}$}{(#2)}}  %
    \else
            \newcommand{\aut}[2]{#1$^{#2}$}
    \fi
    \author{
          \aut{Evyatar Cohen}{1},
          \aut{Jose Yallouz}{1},  
          \aut{Alexander Shpiner}{2},  \\
          \aut{Mark Silberstein}{1},
          \aut{Sylvia Ratnasamy}{3},           
          \aut{Isaac Keslassy}{1,3}
        }%
    \affiliation{
        $^1$ \textit{Technion} \quad 
        $^2$ \textit{Nvidia} \quad 
        $^3$ \textit{UC Berkeley}%
        \country{}%
    }
    \renewcommand{\shortauthors}{E. Cohen \textit{et al.}}     %
\fi

\ifacm %
    \sloppypar
\else 
    \ifhotnets
        \sloppypar
    \else
    \fi
\fi

\begin{abstract}
Mixture of Experts (MoE) architectures have become key to large language models; however, their typical round-robin (RR) scheduling introduces significant bottlenecks. 

In this paper, we demonstrate that RR causes a previously-undiscovered \textit{exponential incast} phenomenon with MoE traffic. We propose an alternative proactive fair scheduling framework tailored for MoE workloads, which effectively prevents fabric oversubscription. We also outline how it can be implemented in NICs.
Finally, through extensive simulations with real and synthetic workloads, we demonstrate that this framework consistently eliminates incast, maintains a near-100\% link utilization, and reduces Collective Completion Time (CCT).
\end{abstract}

\maketitle

\input{01_intro}

\input{02_background}

\input{03_incast}

\input{04_fair}

\input{05_eval}

\input{06_discussion}

\bibliographystyle{ACM-Reference-Format}
\bibliography{mybib} 

\end{document}

%% file: 01_intro.tex
\section{Introduction}
\label{sec:intro}

To overcome the hardware scaling constraints of massive Large Language Models (LLMs), modern data centers are increasingly adopting Mixture of Experts (MoE) architectures. MoE networks scale model capacity by orders of magnitude by dynamically routing individual tokens to a sparse subset of expert layers. However, this application layer efficiency places a severe burden on the underlying network layer. MoE token routing inherently creates highly skewed and fast changing all-to-all communication patterns, which heavily strain datacenter networks~\cite{liu2025netmoe,jiang2024mixtral,jin2025bigmac,gangidi2024rdma,liao2025mixnet,shou2025infinitehbd}.

The MoE pipeline begins with a local routing phase to determine the destination experts for each token, then proceeds in a {dispatch} stage, \ie an all-to-all communication phase where GPUs simultaneously inject token payloads into the fabric. In an environment with $N$ GPU NICs, this dispatch phase causes each node to concurrently flood the fabric with up to $N-1$ variable-length flows. The main goal is to minimize the Collective Communication Time (CCT), defined as the arrival time of the last packet of this dispatch phase at its final receiver.

Scheduling these emerging MoE workloads remains an open challenge. %
Recent works like FAST~\cite{lei2026fast}, Chronos~\cite{renganathan2025chronos} and DFS~\cite{wu2026dynamic} advocate using Birkhoff-von Neumann (BvN) decomposition~\cite{birkhoff1946matrix, vonneumann1953functional} to schedule synchronized, collision-free permutations. However, this tight global synchronization is near impossible to reconcile with distributed Congestion Control Algorithms (CCAs) under variable Round-Trip Times (RTTs), link failures, or background traffic. Consequently, datacenters rely on decentralized heuristics like round-robin (RR), where each NIC independently cycles through its active flows~\cite{nccl,rccl,deepep}. While computationally simple, RR is entirely blind to traffic skewness.

In this paper, we demonstrate that RR suffers from a fundamental breakdown we term the \textit{MoE exponential incast}. When senders independently cycle through asymmetric MoE traffic matrices, they quickly deplete their lighter flows to unpopular experts, leaving only flows for hot GPUs near the end of the epoch. This uncoordinated convergence causes the aggregate arrival rate at bottleneck receivers to spike exponentially. Without a CCA, this incast forces the number of inflight packets to grow exponentially, causing catastrophic buffer overflows. While flows must employ a CCA, such as the emerging Ultra Ethernet Consortium (UEC)
Network-Signaled Congestion Control (NSCC) standard~\cite{uec2025nscc, uec, uec-spec}, to protect switch buffers, we show that such reactive throttling can severely over-correct by throttling too much. This over-correction drops link utilization below 100\%, inflating the CCT, and is increasingly significant as we scale $N$ (\S\ref{sec:evaluation}).

To eliminate this exponential incast, we propose a globally proportional, fair-scheduling framework designed explicitly for MoE workloads. Our core insight is to transform the dynamic scheduling problem into a static, demand-guided rate allocation. By calculating a normalized traffic matrix scaled by the maximum of the worst-case row and column demands, we guarantee that the aggregate arrival rate at any sender or receiver NIC never exceeds its physical link rate. 
Thus, the fabric remains completely collision-free and balanced, preventing the onset of an incast bottleneck by design.
Translating this mathematical framework into a scalable practical system requires sub-microsecond enforcement at the hardware layer. We outline how to achieve this by leveraging for example the NVIDIA DOCA Programmable Congestion Control (PCC)~\cite{nvidia2026doca} runtime integrated directly into modern NICs. By mapping the normalized rate weights directly onto hardware memory lines, our framework bypasses complex network handshakes and control-plane negotiations entirely. We explain that our weight updating scheme can sequentially program up to 1,000 concurrent inter-server flows in under $10\,\mu\text{s}$, an execution window that is practically negligible compared to operational MoE token routing intervals, yielding a scalable network control plane. We also explain why our scheme is resilient to any initial jitter.

We evaluate our proposed fair scheduler across $64 \times 64$ to $512 \times 512$ 
High-radix non-blocking three-tier fat tree, running both a Qwen2/3 MoE workload and a synthetic Zipfian MoE workload. Experimental results firmly corroborate our theoretical framework, showing that our approach consistently outperforms a baseline running RR with UEC's NSCC. 

In summary, this paper makes the following contributions:
\begin{itemize}
    \item We identify and mathematically model the {MoE exponential incast}, a structural bottleneck in decentralized RR scheduling where the MoE skewness causes an exponential  load spike at hot receivers.
    \item We present a proportional rate-allocation framework that normalizes MoE traffic matrices by the worst-case row and column sums, mathematically preventing port oversubscription and incast.
    \item We outline a hardware-enforced architecture utilizing the NVIDIA DOCA Congestion Control interface on modern SmartNICs/DPUs that scales to 1,000 flows in under $10\,\mu\text{s}$.
    \item We evaluate our framework across $64 \times 64$ to $512 \times 512$ network environments, demonstrating a substantial reduction in CCT compared to RR with NSCC.
\end{itemize}

%% file: 02_background.tex
\section{Background and Related Work}
\label{sec:background}

\subsection{Scheduling Goal}
\label{subsec:matrix_representation}
We model the network traffic demand using a demand matrix $D$, 
each entry $D_{ij}$ in the matrix represents the volume of traffic demand originating from sender $i$ and destined for receiver $j$. 
The objective of the scheduler is to compute a transmission sequence that minimizes the {Collective Completion Time (CCT)}, defined as the total time elapsed until all individual flows represented in matrix $D$ have been successfully transmitted across the network fabric.

\subsection{Decomposition-Based Scheduling}
\label{subsec:decomposition_scheduling}
Several papers leverage matrix decomposition techniques to coordinate globally optimal collision-free schedules across the network fabric. 
The FAST framework~\cite{lei2026fast} focuses on achieving optimal scheduling by utilizing the classic Birkhoff-von Neumann (BvN) decomposition theorem~\cite{birkhoff1946matrix, vonneumann1953functional}. Under this approach, the traffic demand matrix $D$ is decomposed into a weighted sum of permutation matrices, ensuring structurally optimal, collision-free fabric allocation. Conversely, the Chronos framework~\cite{renganathan2025chronos} attempts to mitigate the mathematical and computational overhead of exact BvN decomposition by employing a fast, maximal matching decomposition strategy, which sacrifices strict optimality in scheduling efficiency to achieve lower computational complexity.

A key architectural contribution of FAST is that it leverages high-speed intra-server interconnects like NVLink~\cite{nvidia2026nvlink} to handle intra-node communication. By doing so, the framework abstracts away the complexity of GPU-to-GPU routing, allowing the scheduler to focus exclusively on the matrix of transmissions between physical servers. In this work, we leverage this exact same structural paradigm, focusing our optimization solely on inter-server traffic matrices while relying on underlying hardware interconnects to resolve localized intra-node transfers.

However, both FAST and Chronos suffer from several key weaknesses that prevent them from being implemented in practice: (1) While theoretically robust and architecturally streamlined, the primary trade-off of executing an exact Birkhoff-von Neumann (BvN) decomposition in FAST remains its substantial computational complexity, making real-time scheduling a challenge for highly dynamic workloads. (2) The BvN decomposition into permutations does not work when the network has failures, and it cannot be easily adapted to using congestion control.
(3) The BvN decomposition into permutations fits optical switches with the same link delays everywhere and tight timing constraints. But in packet-switched networks, this full complexity of finding permutations is unneeded: (i) Flows have heterogeneous delays depending on the distance between nodes (\eg whether or not the flows cross core switches in fat trees), so all senders and receivers are not fully synchronized, and (ii) we don't need to fully avoid conflicts, as we have packet buffers, we just need to avoid incast.

\subsection{Frameworks}
State-of-the-art distributed LLM frameworks, including NVIDIA's Megatron-LM~\cite{megatron}, ByteDance's veScale/MegaScale~\cite{megascale}, Microsoft's DeepSpeed~\cite{deepspeed}, and DeepSeek's DeepEP~\cite{deepep}.
Starting with local route tokens to target experts at the start of each MoE layer. Before executing the heavy all-to-all dispatch, the participating GPUs must coordinate to prepare receiving memory buffers. 
Crucially, this coordination phase inherently exposes traffic matrix metadata to the devices. 
For instance, in Megatron-LM, GPUs execute a synchronous metadata handshake (collective gather) where each rank exchanges its split sizes, ensuring that every GPU constructs the exact traffic mapping before the physical dispatch begins. 
Similarly, veScale (MegaScale) employs a Single Program, Multiple Data (SPMD) execution model with matching pre-allocated buffers, allowing devices to mathematically derive the traffic matrix without dynamic runtime handshakes. 
Other frameworks only distribute partial mapping information, but 
the full global traffic matrix can be propagated to all participating GPUs with minimal additional control-plane overhead.

\subsection{Additional Related Work}
\label{subsec:related_work_lb}

\mypar{Load-balancing (LB) algorithms} A variety of standard load-balancing (LB) mechanisms can be deployed to distribute MoE traffic across datacenter networks. First, NVIDIA’s Spectrum-X platform~\cite{nvidia2024spectrumx, rizk2023spectrumx} relies on switch-based Adaptive Routing (AR)~\cite{nvidia2023infiniband, abts2022datacenter, voloshin2023congestion} to dynamically route packets based on path congestion. Second, Alibaba implements host-based Oblivious Packet Spraying (OPS)~\cite{lu2025stellar, dixit2013packet, zhang2015lazy, hu2019caps} directly at the endpoints to randomize packet distribution without switch-level coordination. Third, the emerging Ultra Ethernet Consortium (UEC) specification~\cite{uec2025spec} recommends host-based adaptive packet spraying schemes, such as REPS~\cite{bonato2025reps} and STrack~\cite{le2024strack}. In this paper, we evaluate our scheduler across all three routing approaches for completeness.

\mypar{Congestion control algorithms (CCA)} Modern datacenter networks employ various congestion control algorithms (CCAs) to mitigate packet loss and queue buildup caused by transient traffic bursts. In this work, we focus on the recent Network-Signaled Congestion Control (NSCC)~\cite{uec2025nscc, uec, uec-spec} standard, defined by the Ultra Ethernet Consortium (UEC).

%% file: 03_incast.tex
\section{The MoE Exponential Incast}
\label{sec:incast}

\mypar{Intuition}
MoE workloads exhibit highly skewed, dynamic traffic patterns where token routing is dominated by a few highly popular expert servers. Under standard round-robin scheduling, each sender independently cycles through its active flow queues, servicing each active queue at an equal rate. However, because lighter flows to unpopular experts possess significantly fewer tokens, they quickly deplete and drop out of the scheduling pool. Senders are then left exclusively targeting the remaining hot destinations at full line rate, causing a severe network bottleneck that escalates exponentially, a phenomenon we term the \textit{exponential incast}.

To see why this bottleneck surges exponentially, consider the feedback loop inherent to round-robin scheduling. Imagine a sender with one large flow to a hot expert and nine tiny flows to cold experts. Initially, traffic is spread evenly, with only 10\% of the sender's bandwidth going to the hot expert. But as the nine light flows rapidly finish and drop out, the hot flow's share of transmission rate continuously accelerates (from $1/10$ to $1/1$). Because this queue depletion occurs across all senders simultaneously, their remaining transmissions inadvertently synchronize, focusing 100\% of the fabric's capacity onto the hot receiver near the end of the epoch. 

We prove below that the number of remaining destinations decays exponentially fast, and therefore the incast strength also grows exponentially fast. As a result, we also show that nearly the entire workload volume becomes stuck in the network fabric.

\mypar{Theoretical Framework} We model MoE using a Zipfian model and use fluid dynamics to show the resulting impact of RR. Let $N$ be the number of concurrent senders, and let $R = \{r_1, r_2, \dots, r_N\}$ be the set of destination receivers hosting the experts, ordered by their Zipfian popularity rank $j \in \{1, 2, \dots, N\}$, where $r_1$ represents the most requested "hot" expert. We consider the classic Zipfian distribution (with a skewness index $s=1$), where the total traffic volume $D_{ij}$ originating from source $i$ to destination $j$ is given by:
\begin{equation}
D_{ij} = T \cdot \frac{1}{j \cdot \alpha}, \quad \text{with } \alpha = \sum_{m=1}^N \frac{1}{m} \approx \ln N
\end{equation}
and $T$ represents the total traffic volume per sender row. Approximating the harmonic number $\alpha$ as $\ln N$ allows us to ground the fluid dynamics in direct system parameters.

Under RR, each sender loops through its active flow queues independently. If a sender has $k$ active destinations remaining, a round-robin scheduler visits each active queue exactly once per cycle, thereby draining each active flow at an instantaneous rate of $1/k$. 

\T{Exponential decay of active destinations.} Crucially, because lighter flows to less popular receivers ($r_N, r_{N-1}, \dots$) possess significantly fewer tokens, they exhaust their capacity early and drop out of the scheduling pool. To model this in a continuous fluid limit, let $\tau$ represent time. A flow at rank $r$ drains at a rate of $1/k(\tau)$ and is completely depleted at the exact instant $t$ when the boundary of remaining active flows reaches its rank, meaning $r = k(t)$. The accumulated service this boundary flow received up to time $t$ must equal its initial Zipfian volume:
$
\int_0^t \frac{1}{k(\tau)} d\tau = \frac{T}{\alpha k(t)}. 
$
Differentiating both sides with respect to $t$ yields $\frac{1}{k(t)} = -\frac{T}{\alpha k(t)^2} \frac{dk}{dt}$, which we can simplify and solve, yielding
an exponential decay of active destinations:
$
k(t) = N e^{-\frac{\alpha t}{T}} \approx N^{1 - \frac{t}{T}}.
$

\mypar{Exponential incast}
The above decay triggers a severe positive feedback loop for the bottleneck receiver $r_1$. Because $r_1$ remains active throughout the entire epoch, the aggregate arrival rate $\lambda_1(t)$ across all $N$ senders increasingly focuses directly onto it. The incoming load scales inversely with the number of surviving flows, yielding a power-scaling law: $
\lambda_1(t) = N \cdot \frac{1}{k(t)} = e^{\frac{\alpha t}{T}} \approx N^{\frac{t}{T}}.$ 
In other words, initially, traffic is balanced and the hot receiver experiences line-rate capacity ($\lambda_1(0) = 1$). However, as time progresses, the arrival rate surges exponentially until it peaks at $\lambda_1(T) = N$ at the end of the epoch, where all $N$ senders are fully synchronized and transmitting exclusively to $r_1$.

\mypar{Queue dynamics}
To uncover the collective impact of this global incast, we analyze the total network backlog at some time $t = T$. Each sender drains its active queues at a continuous aggregate rate of $k \cdot (1/k) = 1$, introducing exactly $T$ packets into the network over the epoch, which sums to an aggregate injected workload of $NT$. Conversely, assuming each destination receiver services packets at a maximum normalized line rate of $1$, a receiver can clear at most $T$ packets during this period. Due to the severe Zipfian skew, the total arrivals at receiver $j$ are given by $NT / (\alpha j)$. Consequently, receivers can be partitioned into two distinct regimes based on saturation: receiver $j$ is saturated (hot) when $\frac{NT}{\alpha j} > T$ \ie $j < \frac{N}{\alpha}$, and underutilized (cold) otherwise.
While underutilized nodes process all incoming traffic and go idle early, saturated receivers remain bottlenecks for the entire duration, leaving a persistent backlog in their buffers. By integrating the residual queue sizes across all saturated nodes up to the saturation boundary $j_{max} = N/\alpha$, we derive a lower bound on the total number of packets stranded in the network at $t = T$:
$
Q_{total}(T) \approx \int_1^{N/\alpha} \left( \frac{NT}{\alpha x} - T \right) dx = NT \left( 1 - \frac{\ln(\ln N) + 1}{\ln N} \right).
$ 
This closed-form bound reveals an architectural insight for large-scale AI clusters: as $N$ scales out, the second term vanishes, meaning $Q_{total}(T) \rightarrow NT$. \textit{Under RR, nearly 100\% of the entire workload volume becomes completely clogged within the network fabric.} Even though the core bisection bandwidth is theoretically sufficient, independent round-robin selection forces colder destination links to waste valuable service capacity by going idle early, guaranteeing catastrophic cluster-wide buffer oversubscription.

\mypar{Empirical exponential incast}
To validate this exponential incast phenomenon, we conduct a discrete-event network simulation with a real-world $128\times 128$  MoE. 
Senders route their respective traffic matrices using independent, decentralized round-robin scheduling.

\begin{figure}%
    \centering
    \includegraphics[width=0.7\columnwidth]{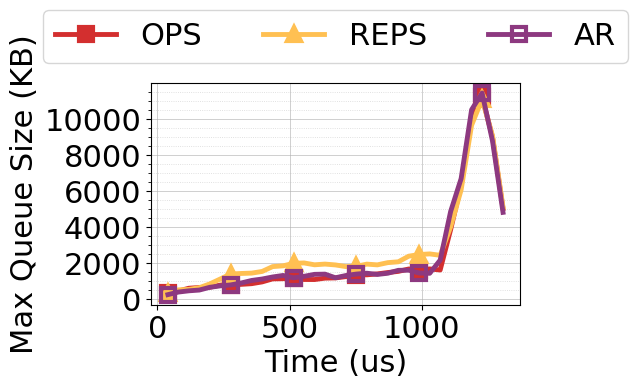}
    \caption{Simulation results showing the queue size at the most requested receiver over time, which grows exponentially near the end of the scheduling epoch.
    }
    \label{fig:exponential_incast_sim}
    \vspace{-10pt} %
\end{figure}
\Cref{fig:exponential_incast_sim} corroborates our fluid model for three different LB schemes  (\cref{subsec:related_work_lb}). Initially, during the first half of the epoch, network utilization remains balanced across the fabric as round-robin flows occupy diverse links. However, as the light and medium-load destination queues are fully depleted, senders progressively synchronize their remaining transmissions toward the heavy-tail expert. After about $t=1,200 \unit{\mu\text{s}}$, in the final $20\%$ of the scheduling timeline, the aggregate ingress rate $\lambda_1(t)$ at the hot receiver exhibits a distinct exponential curve, spiking sharply until the link saturates, triggering massive queue build-ups that match our predicted $Q_{total}(t)$ trajectory and stalling the CCT.

%% file: 04_fair.tex
\section{Fair Scheduling for MoE}
\label{sec:fair_scheduling_moe}

To overcome the exponential incast inherent to uncoordinated round-robin scheduling, we propose a globally proportional, fair scheduling mechanism designed specifically for Mixture of Experts (MoE) workloads. Our core insight is to transform the dynamic scheduling problem into a static rate-allocation framework that mathematically guarantees the prevention of fabric oversubscription.

\mypar{Proportional Rate Allocation}
Our approach relies on computing a normalized traffic matrix $R$ directly from the active demand matrix $D$. Consider a concrete $3 \times 3$ example where three physical servers act as both senders and receivers, and the diagonals are null (signifying that intra-server traffic is offloaded to localized hardware like NVLink):

\begin{equation}
D = \begin{pmatrix}
- & 1000 & 2000 \\
800 & - & 2000 \\
2000 & 1000 & -
\end{pmatrix}
\end{equation}

In this traffic matrix, the third column represents the most heavily loaded bottleneck receiver (expert), accumulating a total incoming demand of $4000$ tokens. To ensure that neither a physical sender's uplink nor a receiver's downlink is oversubscribed, we normalize the demand matrix by a scaling factor $M$, defined as:

\begin{equation}
M = \max \left( \max_{i} \sum_{j} D_{ij}, \ \max_{j} \sum_{i} D_{ij} \right)
\end{equation}

In our specific example, both the worst-case row sum and the worst-case column sum yield a maximum value of $M = 4000$. By dividing the demand matrix $D$ by this bounding factor $M$, we derive the normalized rate matrix $R$:

\begin{equation}
R = \frac{D}{4000} = \begin{pmatrix}
- & 0.25 & 0.5 \\
0.2 & - & 0.5 \\
0.5 & 0.25 & -
\end{pmatrix}
\end{equation}

By construction, this scaling ensures that the sum of every individual row and column in $R$ is strictly less than or equal to $1.0$ ($\sum_{j} R_{ij} \le 1$ and $\sum_{i} R_{ij} \le 1$). If every sender $i$ bounds its transmission rate to receiver $j$ exactly to $R_{ij}$ of the total link line rate, we achieve a fundamental structural guarantee: no individual network port can experience an incast, and fabric utilization remains balanced throughout the entire lifecycle of the MoE epoch.

\section{Implementation Feasibility}

Implementation in a real NIC remains future work. However, we outline below how we plan to do it.

\mypar{Enforcement via NVIDIA DOCA Congestion Control}
To enforce the precisely calculated rate weights $R_{ij}$ directly at the hardware layer, we suggest to leverage the NVIDIA DOCA Congestion Control (DCC) framework integrated into modern BlueField Data Processing Units (DPUs) or ConnectX SmartNICs. 
DOCA Congestion Control provides three distinct architectural advantages that map perfectly to our scheduling goals:
{(1) Fixed hardware rate limiting:} It allows the host to programmatically bind each network flow to a highly granular, fixed transmission weight matching our calculated $R_{ij}$ parameters. 
{(2) Reactive traffic throttling:} In the event of unforeseen transient background traffic, DCC automatically senses link-layer feedback (such as ECN markers) to dynamically scale down rates, providing fallback network stability. 
{(3) Sub-microsecond API updates:} The underlying hardware runtime exposes low-overhead memory mapping registers, making it possible to reconfigure active flow pacing configurations almost instantaneously.

\mypar{Sub-10\,$\mu\text{s}$ scalability}
A common pitfall of global matrix-based scheduling frameworks is the control-plane overhead required to update active network states. To demonstrate the scalability of our approach, we outline a fast runtime weight-updating scheme. 
Our scalability relies on two highly efficient, independent phases. First, finding the bounding factor $M$ for an $N \times N$ matrix ($N \le 1000$) is executed rapidly because the matrix resides in high-speed near-compute memory; consequently, this computation takes under $10\,\mu\text{s}$~\cite{cuda_programming_guide,ampere_microbenchmarking}. Second, writing these calculated rates sequentially to the SmartNIC/DPU hardware memory lines is lightweight and bypasses heavy network handshakes. In addition, instead of writing each of 1,000 rates into a different memory line, we suggest to write all rates together at once into a single memory address. Then, each DOCA PCC flow can access this memory when deciding on its next rate, as already exists today. Writing these $1,000$ rates is expected to complete in under 10 $\mu\text{s}$.
Given that typical MoE token routing intervals operate on significantly longer operational timescales, this configuration latency is practically negligible, demonstrating that our fair-scheduling architecture scales efficiently to large-scale AI data center clusters.

%% file: 05_eval.tex
\section{Evaluation}
\label{sec:evaluation}

\mypar{Simulator and Topology}
We simulate our framework using htsim~\cite{htsim-fork}, chosen for its ability to scale to large datacenter networks and its existing implementation of transport and load-balancing protocols~\cite{gerstein2025}. Specifically, we model a high-radix, non-blocking leaf-spine network fabric with $800\unit{Gbps}$ links. Senders transmit packets with payload size of $4,096$ bytes. As in FAST, we assume that groups of 8 physical servers leverage local NVLink for intra-node communication, yielding a modified traffic matrix that focuses only on traffic that doesn't go through NVLink. We benchmark cluster scales from $64 \times 64$ to $512 \times 512$ GPUs.

\mypar{Load-balancing} The simulator models three LB schemes: AR, OPS, and REPS (\cref{subsec:related_work_lb}).

\mypar{Workloads} To capture the precise traffic dynamics of modern Mixture of Experts (MoE) infrastructures, our evaluation relies on two distinct dataset paradigms:

\mypars{Real-world MoE routing workloads:} To anchor our framework in true production conditions, we extract trace datasets collected directly from the active routing phase of a production-grade MoE. They map the actual token-to-expert routing decisions, capturing real-world non-uniformities for $64 \times 64$ (Qwen2) and $128 \times 128$ matrices (Qwen3-235B).

\mypars{DeepSeek-V3 parameterized synthetic workloads:} We generated synthetic dense matrices skewed via a Zipfian distribution ($s = 0.25, 0.5$), replicating scenarios where tokens are dynamically routed to 1, 2, 4, or 8 experts. Token counts and payload sizes are parameterized directly from the DeepSeek-V3 architecture~\cite{deepep}.

\mypar{Baselines}
We evaluate our proposed rate-based scheduler against a baseline denoted \textit{Round-Robin with Congestion Control} (RR-NSCC) using the \texttt{htsim} simulator. The core difference lies in how packets are queued and scheduled at the host: (1) under our proposed approach, each flow is assigned a specific, calculated transmission rate that dictates its pacing; (2) under the RR-NSCC baseline, flows are allowed to inject up to two packets into a shared NIC queue. Then, this shared queue employs a fairness mechanism that cyclically selects and transmits a single packet from each active flow in a round-robin order (default htsim behavior).

\begin{figure}[t]
    \centering
    \includegraphics[trim={0pt} {30pt} {0pt} {0pt}, clip, width=0.9\columnwidth]{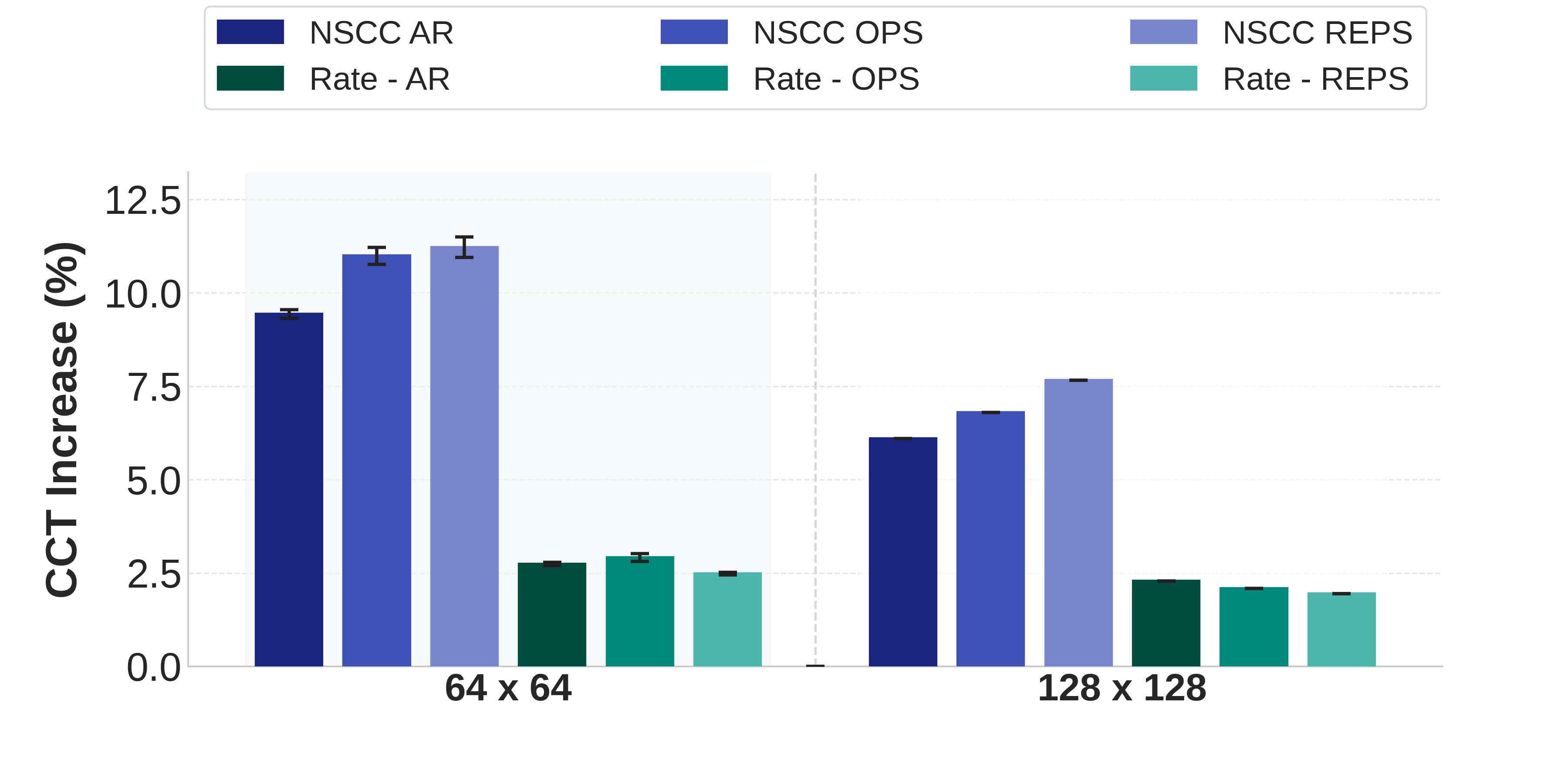}
    \caption{Comparison of the CCT inflation relative to a theoretical lower bound (assuming a constant 100\% link utilization) under empirical MoE LLM routing traces across $64 \times 64$ (left) and $128 \times 128$ (right) GPUs matrices.}
    \label{fig:cct_comparison_llm}
        \vspace{-10pt} %
\end{figure}

\begin{figure*}
    \centering
    \includegraphics[trim={0pt} {30pt} {0pt} {20pt}, clip, width=0.8\textwidth]{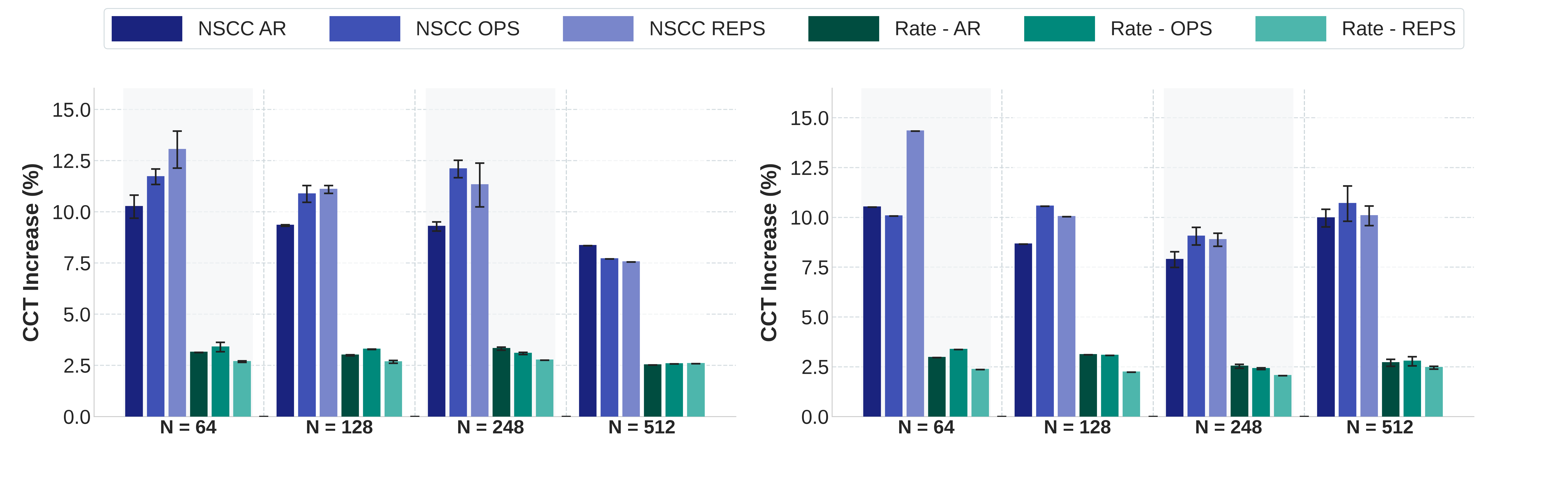}
    \caption{CCT inflation compared to an ideal lower bound (constant 100\% link utilization) under heavily skewed Zipf distributions $s=0.25$ (left) and $s=0.5$ (right), evaluated across OPS, REPS, and AR routing configurations.}
    \label{fig:cct_comparison_zipf}
        \vspace{-3pt}
\end{figure*}

\begin{figure}%
    \centering
    \includegraphics[width=0.6 \columnwidth]{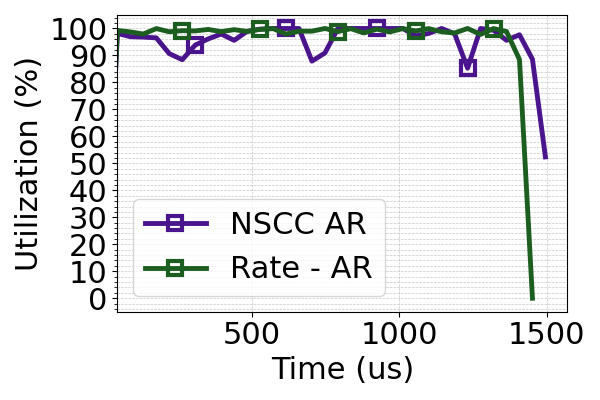} 
    \caption{The queue-utilization trade-off: reactive NSCC over-corrects, dropping link utilization below 100\% during active transmission and inflating CCT. Rate-based scheduling maintains full utilization.
    }
    \label{fig:nscc_tradeoff}
        \vspace{-10pt} %
\end{figure}

\mypar{Collected workload}
The macro evaluation results demonstrate a distinct advantage for our proposed rate-based fair scheduling framework in reducing the overall workload completion time. Below, we analyze these performance gains across both real-world  and synthetic traffic matrices, while illustrating the underlying network-level trade-offs.

\mypar{Real-world MoE traces}
We first evaluate our framework using real-world MoE traces. Specifically, we analyze approximately 80 real-world MoE routing traffic matrices of size $64 \times 64$. As illustrated in \cref{fig:cct_comparison_llm} (left), our rate-based approach yields a pronounced and consistent reduction in CCT compared to the state-of-the-art RR-NSCC baseline. 
When scaling the network size to $128 \times 128$ (\cref{fig:cct_comparison_llm}, right), our framework continues to outperform RR-NSCC, though the relative performance improvement is slightly less pronounced than in the $64 \times 64$ scenarios. This marginal narrowing of the gap is primarily attributed to a limitation in our empirical dataset; we had access to only a single, isolated traffic matrix for the $128 \times 128$ scale, making this specific data point less statistically representative of the broader workload distribution. Nonetheless, the consistent advantage highlights the robustness of our approach on real-world workloads.

\mypar{Sensitivity to skewness and number of nodes}
To rigorously assess our framework under extreme and varied traffic conditions, we evaluated its performance using synthetic traffic matrices modeled after DeepSeek-style MoE patterns. As shown in \cref{fig:cct_comparison_zipf}, our rate-based scheduler delivers a clear and systematic advantage across all evaluated network sizes (ranging from $64 \times 64$ up to $512 \times 512$ nodes) and across all tested Zipf skewness factors ($s=0.5$ and $s=0.25$). Even under severe skewness, our balanced allocation prevents the exponential incast that cripples standard Round-Robin scheduling, demonstrating that the theoretical design scales robustly with the network fabric.

\mypar{The queue-utilization trade-off}
To validate the physical mechanism driving these performance gains, we analyzed the link utilization over time using the real-world MoE $64 \times 64$ routing matrix. The results in \cref{fig:nscc_tradeoff} strongly corroborate our core theoretical premise. While the reactive NSCC baseline successfully prevents catastrophic buffer overflows by capping queue occupancy, it does so at a steep cost: the reactive throttling heavily over-corrects, dropping link utilization significantly below 100\% for extensive intervals during the active simulation phase. This transient starvation of the fabric's bandwidth directly inflates the overall CCT. In sharp contrast, our rate-based scheduling framework maintains a near-perfect, consistently high utilization profile throughout the active transmission window, proving that proactive, fair rate computation eliminates the need for destructive, reactive throttling cycles.

%% file: 06_discussion.tex
\section{Impact on Future Network Architectures}
\label{sec:discussion}

The shift from reactive, heuristic-based network scheduling to a proactive, proportional rate-allocation framework introduces several significant changes for large-scale AI fabrics.

\mypar{Redefining the role of transport-layer congestion control} 
Historically, data center networking has relied heavily on complex reactive transport-layer protocols (e.g., DCQCN, HPCC, or the emerging UEC NSCC standard)  to handle traffic bursts. Our method suggests a paradigm shift: by pre-normalizing the traffic matrix at the application layer relative to the physical capacity of the bottlenecks, the network fabric becomes inherently collision-free. This could significantly simplify the need for aggressive, fine-grained hardware congestion loops, reducing the computational overhead and silicon real estate devoted to packet-throttling logic on future NICs and switches.

\mypar{Enabling asymmetric and heterogeneous AI clusters} 
Current distributed platforms typically assume homogeneous hardware clusters (identical servers and NICs) to prevent slower paths from stalling the global synchronization barrier. Because our normalization scaling factor $M$ dynamically scales by worst-case demands, the algorithm can easily adapt to asymmetric NIC capacities. Future clusters could safely mix older 200 Gbps servers with newer 400 Gbps/800 Gbps server models, or feature unbalanced expert placements; our scheduling matrix would naturally compute optimal proportional rates that ensure the slower hardware never creates an accidental incast bottleneck.

\mypar{Mitigating the cost and tail-latency of buffer bloat} 
As network speeds scale to 1,600 Gbps and beyond, building high-speed on-chip switch buffers becomes prohibitively expensive and physically limited by silicon scaling. The exponential incast highlighted above demonstrates that traditional scheduling fills these buffers rapidly, leading to tail-latency degradation. By pacing flows proportionally so that the normalized ingress rate never exceeds 1, our algorithm significantly reduces switch queue depths. This could allow future data center switches to be designed with significantly smaller buffers, driving down the overall capital expenditure and power envelope of next-generation network hardware.